\documentclass[10pt,conference]{IEEEtran}
\IEEEoverridecommandlockouts
\usepackage{cite}
\usepackage{amsmath,amssymb,amsfonts}
\usepackage{algorithmic}
\usepackage{graphicx}
\usepackage{textcomp}
\usepackage{xcolor}
\usepackage{hyperref}
\usepackage{subcaption}

\def\BibTeX{{\rm B\kern-.05em{\sc i\kern-.025em b}\kern-.08em
    T\kern-.1667em\lower.7ex\hbox{E}\kern-.125emX}}

\begin{document}

\title{DroneWorld: High Fidelity Simulation tool for testing small Uncrewed Aerial System Applications}
\title{DroneReqValidator: Facilitating High Fidelity Simulation Testing for Uncrewed Aerial Systems Developers}
\author{\IEEEauthorblockN
{Bohan Zhang, Yashaswini Shivalingaiah, and Ankit Agrawal }
{bohan.zhang.1@slu.edu, yashaswini.shivalingaiah@slu.edu, ankit.agrawal.1@slu.edu}
\IEEEauthorblockA{\textit{Department of Computer Science, Saint Louis University} \\
Saint Louis, MO, USA }
}

\maketitle

\begin{abstract}
Rigorous testing of small Uncrewed Aerial Systems (sUAS) is crucial to ensure their safe and reliable deployment in the real world. sUAS developers aim to validate the reliability and safety of their applications through simulation testing. However, the dynamic nature of the real-world environment, including factors such as challenging weather conditions and wireless interference, causes unique software faults that may only be revealed through field testing. Considering the high cost and impracticality of conducting field testing in thousands of environmental contexts and conditions, there exists a pressing need to develop automated techniques that can generate high-fidelity, realistic environments enabling sUAS developers to deploy their applications and conduct thorough simulation testing in close-to-reality environmental conditions. To address this need, DroneReqValidator (DRV) offers a comprehensive small Unmanned Aerial Vehicle (sUAV) simulation ecosystem that automatically generates realistic environments based on developer-specified constraints, monitors sUAV activities against predefined safety parameters, and generates detailed acceptance test reports for effective debugging and analysis of sUAV applications. Providing these capabilities, DRV offers a valuable solution for enhancing the testing and development process of sUAS. The comprehensive demo of DRV is available at 
\url{https://www.youtube.com/watch?v=Fd9ft55gbO8}


\end{abstract}


\section{Introduction}


The advancement of artificial intelligence has led to the evolution of smart small Uncrewed Aerial Systems (sUAS). The applications of smart sUAS have expanded from package delivery \cite{chen2022dronetalk} in densely populated cities to military reconnaissance in remote areas \cite{stodola2019cooperative}, incorporating advanced decision-making capabilities to enable autonomous missions in various environmental conditions~\cite{ilachinski2017artificial}. These novel applications of Unmanned Aerial Vehicles (UAVs) have posed numerous challenges in testing their safety and reliability in adverse environmental contexts such as wind gusts, non-uniform deterioration of satellite signals, and fluctuations in cellular or 5G signals.

During system validation, UAV software application developers encounter two primary issues. Firstly, while simulation testing helps in validating the behavior of the UAV application, it often fails to accurately represent the realistic and dynamic nature of the operational environment, resulting in failures during field deployments. Further, certain field tests present substantial difficulties when performed on physical UAVs, especially those that involve pushing or exceeding operational limits, such as flying in extreme weather conditions or close proximity to objects or humans. Secondly, UAV application developers are usually experts in Software Engineering, while developing highly realistic 3D environments requires expertise in graphics, design, and specialized tools such as Unity \cite{UnityRea43:online} or Unreal Engine \cite{unrealengine}. This gap in skills presents a major challenge for sUAS developers in their ability to efficiently and rapidly generate the desired 3D environment for their simulation testing needs.


\begin{figure}[t]
    \centering
    \includegraphics[width=\columnwidth]{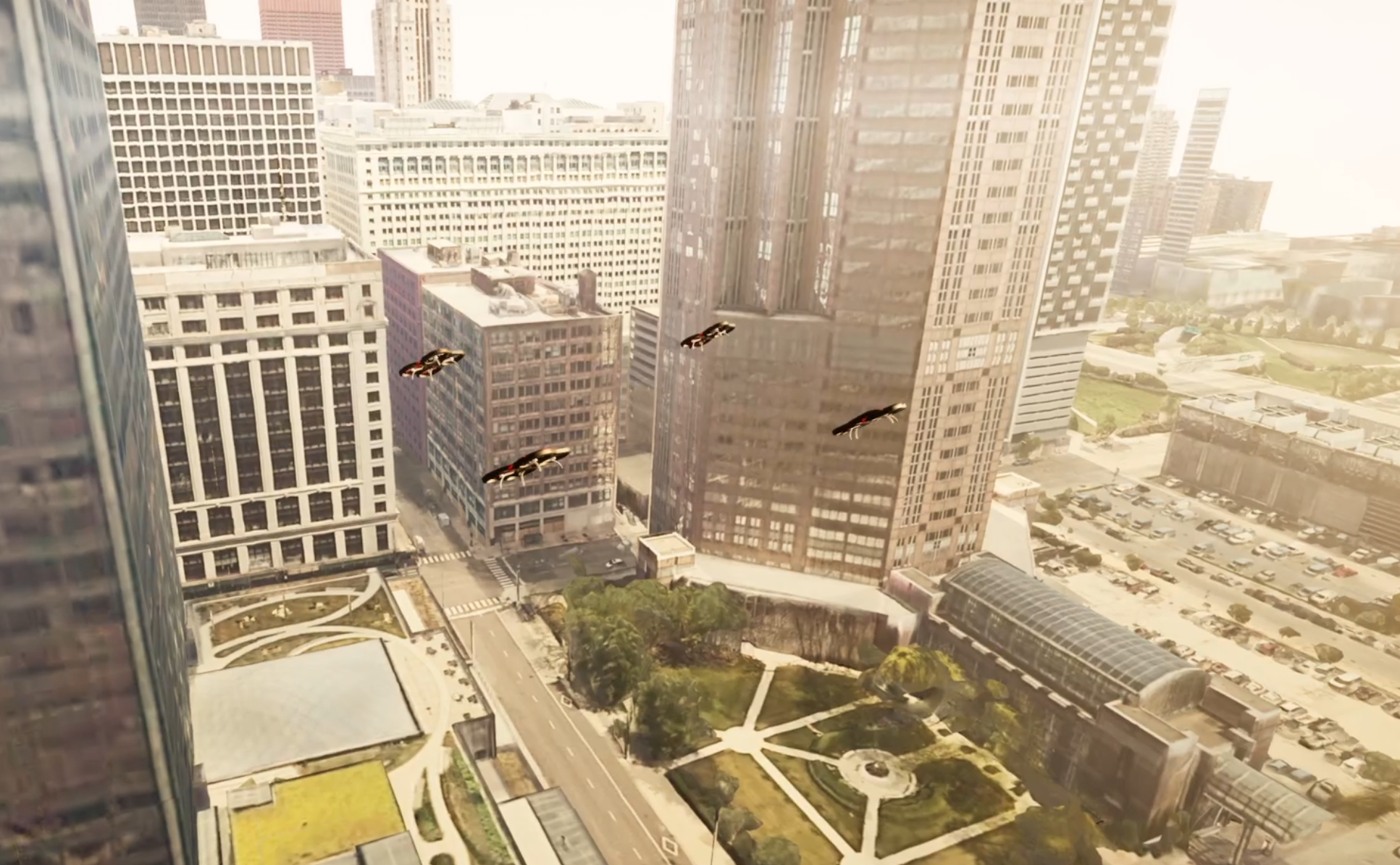}
    \caption{Four UAVs surveilling the realistic digital twin model of Chicago city in DroneReqValidator}
    \label{fig:Chicago4Drone}
    \vspace{-20pt}
\end{figure}
Currently, sUAS developers heavily rely on simulations using either 2D maps or 3D simulation environments like Gazebo \cite{koenig2004design} or AirSim \cite{shah2018airsim}. However, these existing simulation tools suffer from certain limitations. First, they require the manual construction of 3D environments, increasing developers' workload. This manual process is time-consuming, resource-intensive, and can introduce inaccuracies and limitations in the realism of the simulated environments. Second, the responsibility of collecting and analyzing simulation data falls on the developers as they need to design and implement custom code for relevant data collection, plotting, and analysis. The lack of automated data handling tools adds to the complexity of the testing process and further increases the developers' workload and reduces productivity \cite{murphy2019predicts,cheng2022improves}. Third, these simulation tools lack a standardized mechanism for specifying and evaluating test conditions ~\cite{zheng2015perceptions}. Due to this limitation, testing through the existing simulation tool results in ad-hoc simulation testing. In summary, current UAV simulation tools do not offer a complete testing harness that can be effectively used by large teams of sUAS developers.

In contrast, our tool, DroneReqValidator, referred to as DRV in the remainder of the paper, offers a comprehensive simulation ecosystem for sUAS simulation testing. The primary goal of DRV is to automate the generation of realistic environments, based on developer-specified constraints, and facilitate the monitoring of sUAV activities against predefined safety properties. Additionally, DRV generates detailed acceptance test reports containing graphical plots and clear descriptions of failed and passed test cases, which are useful for effective debugging and analysis of sUAS. DRV leverages the advancements in Unreal Engine \cite{unrealengine}, Google Earth digital twin models \cite{googlemaps2021tile},  and AirSim APIs\cite{shah2018airsim} to enable the automatic generation of a realistic simulation environment and simulating multiple UAVs in them. DRV's ultimate goal is to provide sUAS application developers with a specialized test harness that serves as a structured framework for simulation testing, facilitate simulation testing in large-scale sUAS projects, assist developers in identifying problems and faults resulting from the dynamic nature of the environment during simulation testing, and increase developer confidence before field testing. 


The remainder of this tool demo paper is organized as follows. Section \ref{sec:arch} describes DRV's architecture, and discusses its main components and features, while Section \ref{sec:challenges} outlines preliminary analysis, and open challenges, and discusses DRV's current status. Finally, Section \ref{sec:code} includes links to the codebase of every component of DRV.

\section{DRV Architecture}
\label{sec:arch}

DRV employs a client-server architecture consisting of a front-end and a back-end to promote collaboration among developers during simulation testing. 
The front-end allows developers to configure simulation environments based on the application requirements and acceptance tests, while the back-end executes simulations in the digital twin models of specified geographical locations and generates acceptance test reports. The following section provides details of each component.  

\subsection{Front-End Client}
\label{sec:front-end}
DRV's front-end component is a user-friendly web application built with the popular React web framework. It offers an intuitive Graphical User Interface (GUI) that simplifies the process of setting up the simulation environment, specifying UAV sensor configuration, and configuring safety test properties. DRV uses a three-step wizard interface to guide users seamlessly through the configuration of each aspect of simulation testing.


\subsubsection{\textbf{Realistic Environment Configuration}}
The first step in the wizard enables developers to specify the key elements of a realistic simulation environment. These elements include the geographical region in which developers wish to conduct the simulations, weather conditions including wind, and the specific time of day. Based on these inputs, the system's back-end creates a 3D digital twin model of the specified region, incorporates wind in the designated direction and velocity, and adjusts lighting conditions according to the specified time of day. The details of the back-end are discussed in Section \ref{sec:backend}. Figure \ref{fig:Chicago4Drone} shows four UAVs flying in DRV's digital twin model of Chicago in the afternoon.



\subsubsection{\textbf{UAV Sensor Configuration}}
Multi-UAV software applications could include heterogeneous UAVs and therefore developers need the ability to easily configure the sensor properties of each participating UAV in the mission and conduct UAV-specific acceptance testing. DRV's GUI caters to this requirement by allowing developers to configure the specifications of each UAV involved in the mission. This includes defining the number of sUAVs to be deployed, specifying the sensor configurations for each individual sUAV, and setting their respective home geolocations within the simulation environment. The current version of DRV allows developers to configure major sensors such as GPS, Camera, Lidar, Barmoter, and Magnetometer. 
Based on these inputs, the back-end deploys multiple sUAS with the defined sensor configurations at the specified geo-locations in the simulated environment. To further explore the user interface and available configurable properties, please refer to the DRV Demo video. 

\begin{figure}[htbp]
    \includegraphics[width=\columnwidth]{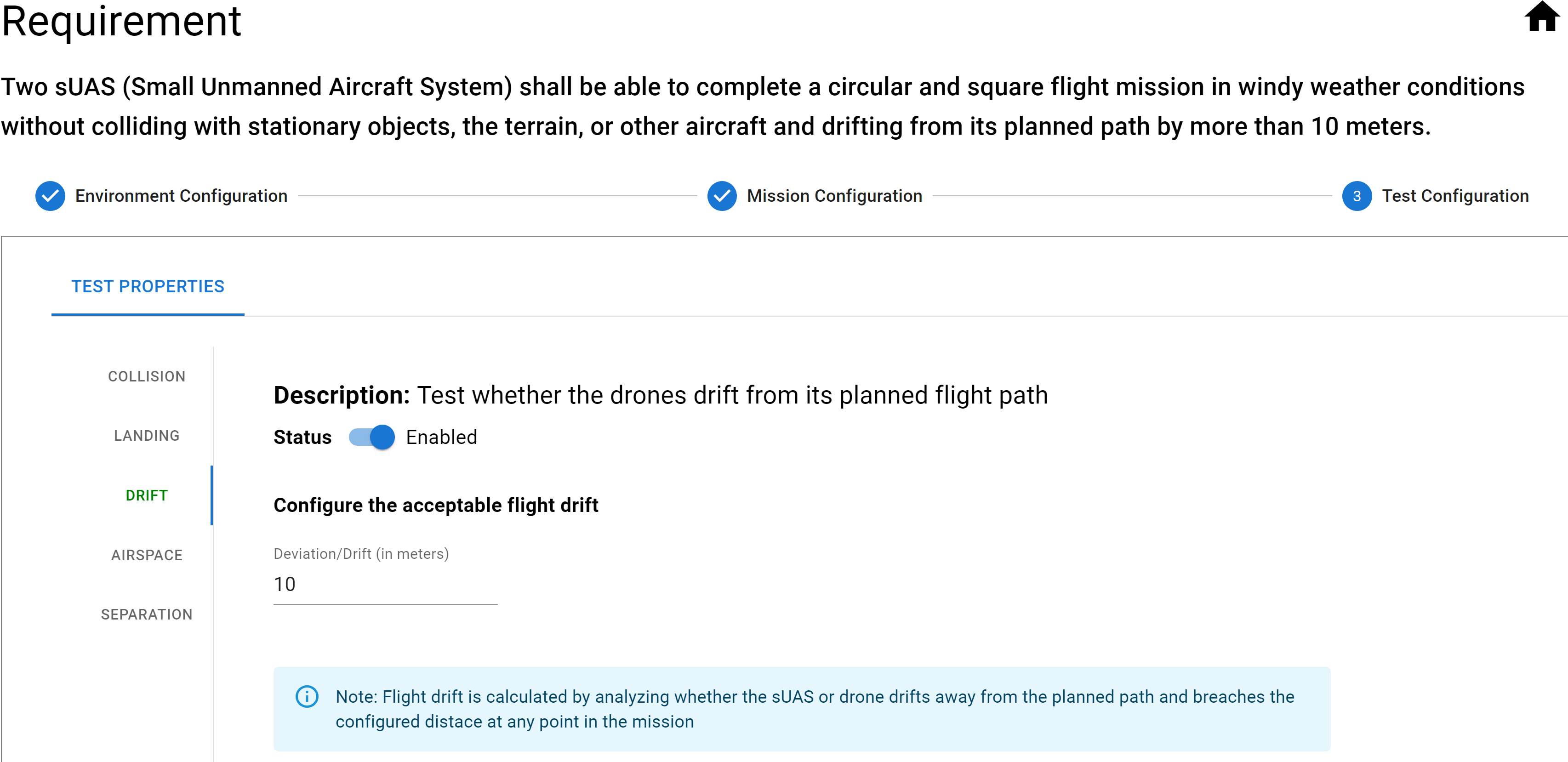}
    \caption{DRV UI to specify the safety test properties }
    \label{fig:testProperty}
    \vspace{-8pt}
\end{figure}

\subsubsection{\textbf{Test Property Configuration}}
One of the primary objectives behind the development of DRV is to facilitate the seamless specification of pass and fail criteria during simulation testing. As a crucial step in the configuration, users are required to specify a range of metrics and conditions that determine the success of a simulation. Figure \ref{fig:testProperty} demonstrates the user interface for configuring test properties. In our current version of DRV, developers have the ability to define test cases based on various metrics. These metrics include configuring the maximum permissible deviation from the intended flight path, configuring the minimum separation distance UAVs shall maintain from each other, tracking the occurrences of collisions with environmental elements, configuring safe landing spots and enabling tracking UAV landings, configuring no-fly zones and tracking whether UAVs enter such restricted zones or not during the simulation. These preliminary test properties were specifically selected based on sUAS hazards and safety requirements \cite{agrawal2019leveraging, vierhauser2019interlocking}.  


\begin{figure}[htbp]
    \centering
    \includegraphics[width=\columnwidth]{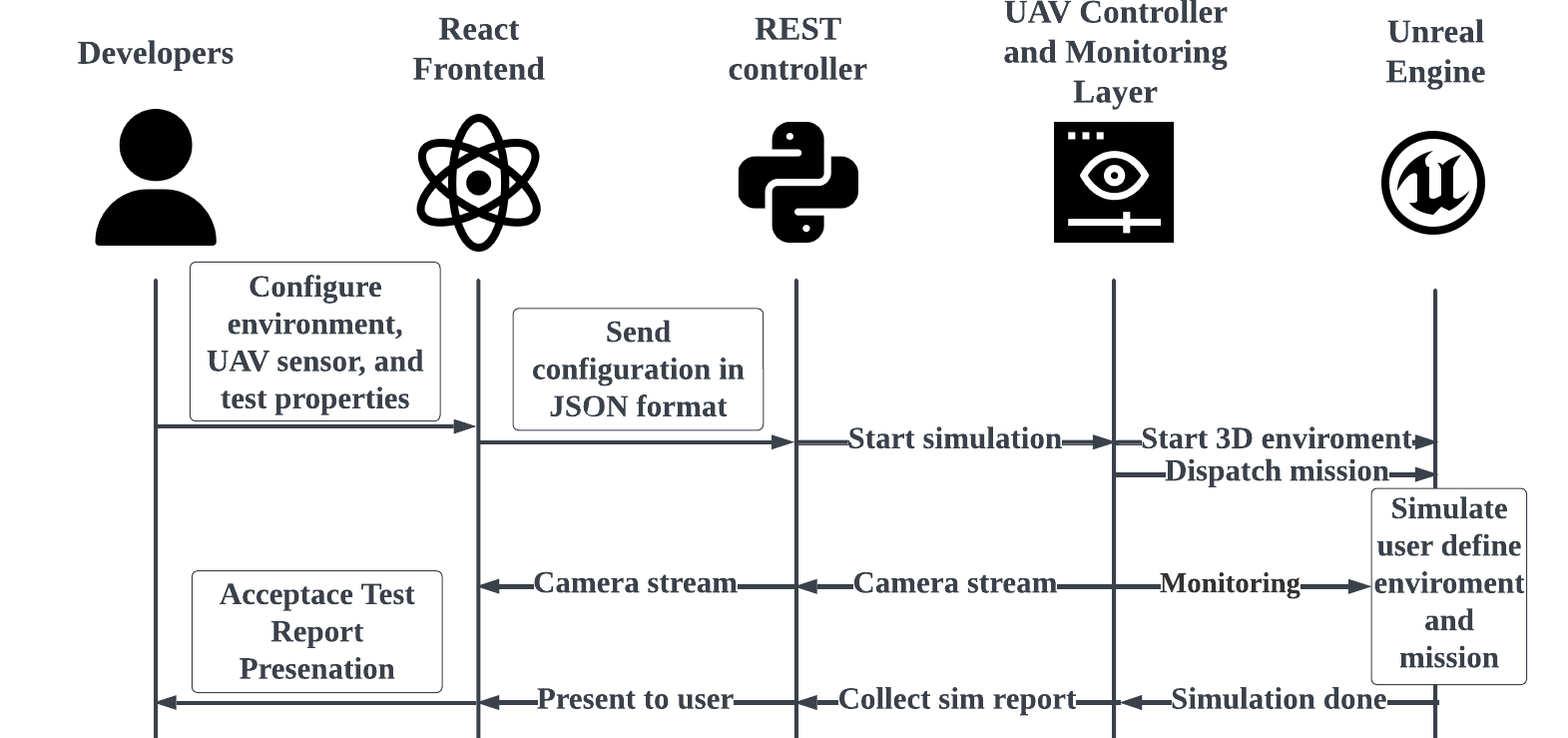}
    \caption{Flow Chart depicting the interaction between major components of DRV}
    \vspace{-20pt}
    \label{fig:dwarch}
    
\end{figure}

\subsection{Back-End Server}
\label{sec:backend}
The back-end server of DRV is built entirely in Python and utilizes the Flask framework. It comprises of several key components, including the REST Controller, UAV Controller, and Monitoring layers. The REST controller layer is primarily responsible for receiving and processing the developers' specifications regarding the environment, UAV sensors, and the safety test properties provided through the front-end client. The REST controller layer also validates the input data and transfers it to the control and monitoring layer for further processing.

 The UAV Controller and Monitoring layer establish communication with Unreal Engine \cite{unrealengine} via Airsim APIs \cite{shah2018airsim}, allowing the back-end to instantiate the 3D environment and accurately simulate the behavior of UAVs within it. UAV Controller and Monitor manage tasks through a task queue, ensuring that developers' simulation requests are processed in the order they are received. This approach enables effective control and monitoring of multiple UAVs in the simulation environment while preserving the integrity of the overall simulation process. The interaction between the different components is shown in Figure \ref{fig:dwarch}. The unique features of the simulation are discussed next.


\subsubsection{\textbf{Automatic Realistic Environment Generation}}DRV's Automatic Environment Generation component utilizes the advanced capabilities of the Unreal Engine to deliver highly realistic 3D environments, as depicted in Fig. \ref{fig:Chicago4Drone}. 

DRV's current version encompasses four distinctive maps for diverse testing: \textit{Blocks}, \textit{City Park}, \textit{Chicago Scanning}, and \textit{Cesium}. The \textit{Blocks} map offers a simplified and lightweight 3D environment, suitable for quick testing of sUAS functionality. The \textit{City Park} map showcases the Unreal Engine's cutting-edge technologies such as Lumen lighting and Nanite, enabling highly realistic urban environment simulations. The \textit{Chicago Scanning} map provides a precise 3D representation of the city of Chicago, serving as an ideal testbed for evaluating sUAV maneuverability in dense urban settings. Finally, the \textit{Cesium} map supports real-time streaming of satellite scanning and terrain data of real-world geolocations, enabling developers to simulate sUAS in real-world locations with accurate terrain representation.


\subsubsection{\textbf{Automated Fuzzy Testing}}

Fuzzy testing \cite{nguyen2022bedivfuzz} is a software testing technique that aims to discover vulnerabilities in the system's behavior by changing the inputs to the system iteratively. In sUAS applications, where environmental conditions can be unpredictable and challenging, fuzzy testing becomes particularly valuable. DRV automatically manipulates the developers' specified environmental factors to test UAV application behavior in extreme and diverse environmental conditions. This automated testing approach enables developers to determine the system's performance limits in real-world scenarios.

\subsubsection{\textbf{Runtime Monitoring and  Report Generation}}
\begin{figure}[htbp]
    \centering
    \includegraphics[width=\columnwidth]{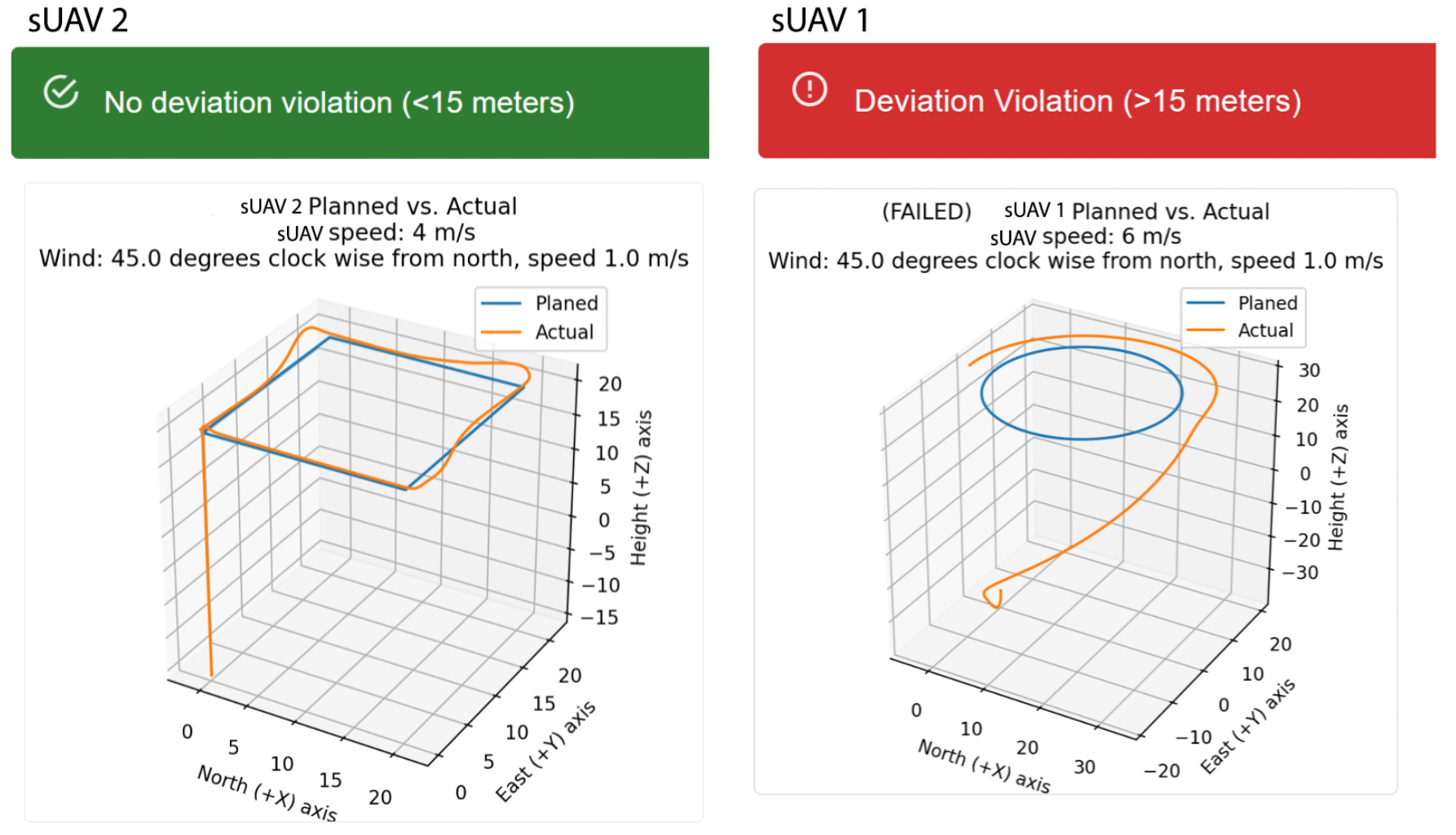}
    \caption{Report snippet showing pass/fail acceptance tests}
    \label{fig:acceptance}
    \vspace{-10pt}
\end{figure}
 
Runtime monitors collect data and observe the interaction between the system and the environment \cite{rabiser2017comparison}. DRV utilizes AirSim APIs to track the activities of simulated UAVs in the 3D environment. The runtime monitors continuously collect sensor data from the UAVs, including position, speed, altitude, camera feed, lidars, and telemetry. This data enables identifying any safety property violations in the UAV behavior. When a safety violation is detected, potential issues are logged and made available to the report generator. The details of the violations, along with data analysis in the form of graphical plots are then rendered on the user interface for developers to analyze.

Fig. \ref{fig:acceptance} presents two instances of the graphical plots in the acceptance report for the UAV's maximum flight path deviation safety property, highlighting the successful case in green and the failed case in red. This visual representation offers additional insights and specificity to the developers.

These automatically generated reports are a valuable resource for sUAS application developers as they streamline the process of identifying and diagnosing issues, saving time and effort in development and troubleshooting. Additionally, the automatically generated acceptance test report can be augmented as evidence in Safety Assurance Cases to argue the safety of the sUAS \cite{agrawal2019leveraging,vierhauser2021hazard}.

\section{Ongoing Development and Challenges}
\label{sec:challenges}
\begin{itemize}
\item 
\textbf{Simulation at Scale}: As the number of Unmanned Aerial Vehicles (UAVs) in the simulation increases, the demand on both the Central Processing Unit (CPU) and Graphical Processing Unit (GPU) also increases. To illustrate this challenge, a stress test was conducted on a machine with the specifications outlined in Table \ref{tab:sys_specs}. The findings of the stress test are depicted in Figure \ref{fig:5drone}, which illustrates the CPU and GPU usage over time during a simulation involving 5 and 20 sUAVs, respectively. First, the graph shows that CPU demand increases as the number of UAVs in the simulation environment increases. The CPU demand increases because the UAV Controller and the Monitors component rapidly switch threads, impacting performance. Second, interestingly, we also observed that the GPU utilization for a fleet of 20 sUAVs  is comparatively low. In this case, the CPU bottleneck hinders the rendering of the environment, thereby decreasing the demand for GPUs.

\begin{table}[htbp]
    \centering
    \begin{tabular}{|c|l|}
    \hline
        \textbf{Component} & \textbf{Specification} \\ \hline
        \textbf{GPU} & NVIDIA GeForce RTX 3090 Ti 24G \\ \hline
        \textbf{CPU} & Intel Core i9-10900 \\ \hline
         \textbf{RAM}& 64 GB \\ \hline
    \end{tabular}
    \caption{System Specifications for Preliminary Stress Tests}
    \label{tab:sys_specs}
    \vspace{-25pt}
\end{table}

\begin{figure}[hbt!]
    \centering
    \includegraphics[width=0.8\columnwidth]{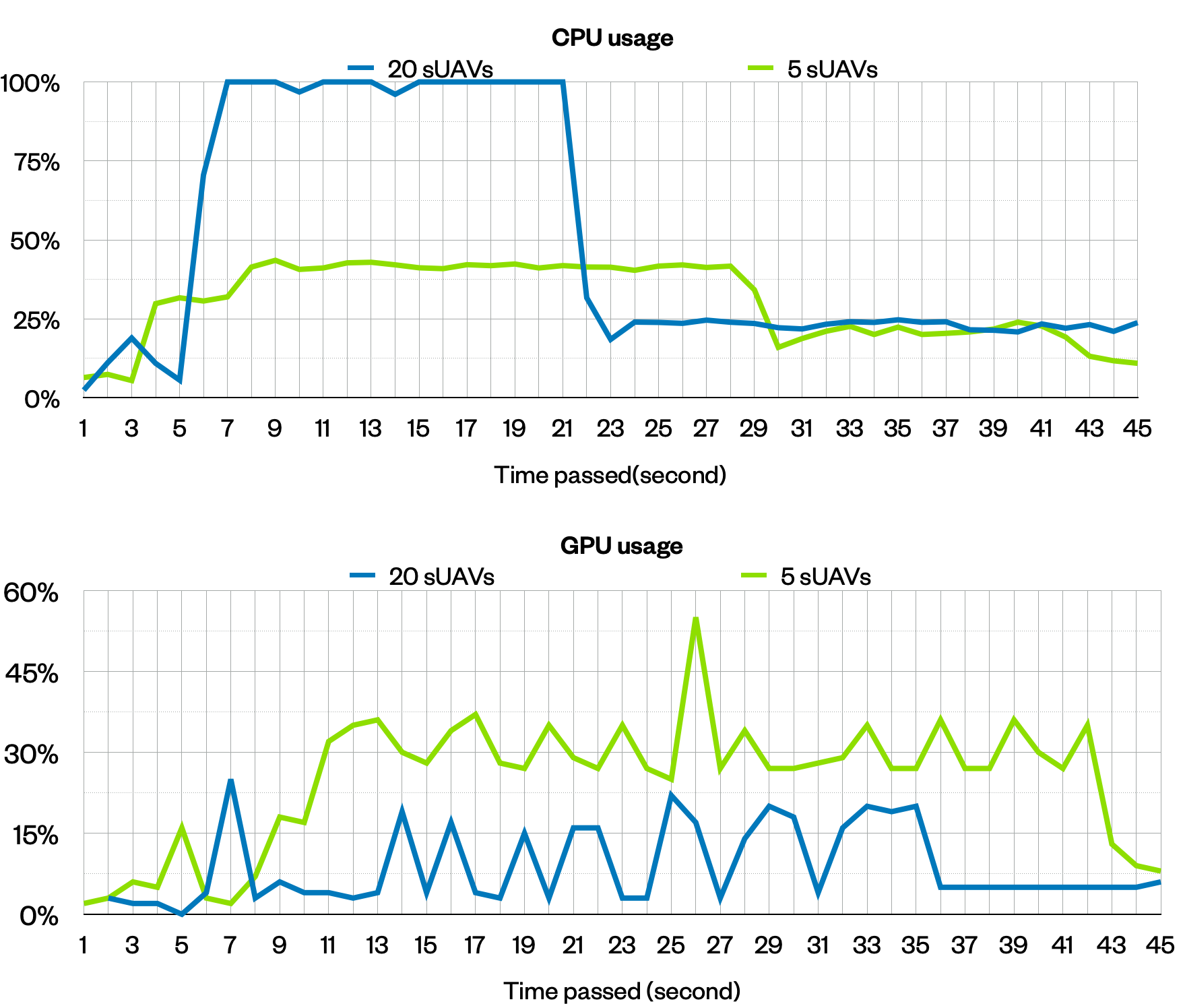}
    \caption{Stress Test Results: 5 Vs 20 sUAVs simulation}
    \label{fig:5drone}
    \vspace{-10pt}
\end{figure}

\item \textbf{Geo-Cordinate System Alignment}: 

DRV integrates the AirSim and Cesium Unreal plugins along with Google Earth APIs in the Unreal Engine to simulate UAVs in realistic geographical environments. However, inconsistencies in the coordinate systems between AirSim and Cesium pose challenges in accurately aligning geo-coordinates and elevation data. During our initial testing phase, we observed a minor discrepancy between the physical movement of a sUAV in the real world and its corresponding translation within the 3D environment of DRV. Our testing also revealed that Google Maps elevation data occasionally does not match the elevation in the Cesium, resulting in UAVs flying below terrain unintentionally. By implementing precise coordinate transformation algorithms, calibrating the scaling factors, and validating elevation data, we are actively addressing the challenges associated with coordinate alignment and elevation.


\item \textbf{Continious Integration Pipeline Integration}: DRV currently operates as a client-server application, but we're actively working on implementing a Continuous Integration (CI) pipeline for enhanced testing. CI pipeline will automatically execute simulation tests in DRV after every commit to the sUAS application's Git repository, reducing developers' manual testing workload. Moreover, the availability of acceptance test reports after each commit will enable quick identification of issues resulting from specific commits to the sUAS applications's code repository.



\end{itemize}
\section{CODE AVAILABILITY}
\label{sec:code}

All components of the DRV ecosystem are available on our public Github respository as open source projects \footnote{\url{https://github.com/UAVLab-SLU/DRV_public}}.










\bibliographystyle{abbrv}
\bibliography{droneSim}

\end{document}